\documentstyle[12pt]{article}
\advance\textheight by \topskip
\begin{document}
\input epsf

\thispagestyle{empty}
\begin{flushright}
{SU-ITP-96-01} \\ astro-ph/9601004 \\ January 1, 1996
\end{flushright}
\begin{center}
{\Large\bf RECENT PROGRESS\\
\vskip .5cm
 IN INFLATIONARY COSMOLOGY\footnote{An invited talk at the  1st RESCEU
International Symposium on "The
Cosmological Constant and the Evolution of the Universe," Tokyo, November
1995.}}
\vskip 1.1cm {\bf
Andrei Linde} \\
\vskip
0.3cm Department of Physics, Stanford University, Stanford CA
94305-4060, USA
\end{center}

\vskip .5 cm {\centerline{\large ABSTRACT}}
\begin{quotation} \vskip -0.3cm
We discuss two important modifications of inflationary paradigm. Until very
recently we  believed that inflation automatically leads to flatness of the
universe, $\Omega = 1\pm 10^{-4}$.  We also thought   that post-inflationary
phase transitions in GUTs may occur only after thermalization, which made it
very difficult to have baryogenesis in GUTs and to obtain superheavy
topological defects after inflation. We  will describe a very simple version of
chaotic inflation  which  leads to a
division of the universe into infinitely many open  universes with all possible
values of $\Omega$ from $1$ to $0$.  We will show also that in many
inflationary models quantum
fluctuations of  scalar and vector fields produced during reheating
are much greater than they  would be
in a state of thermal equilibrium. This   leads to   cosmological
phase transitions of a new type, which may result in an efficient GUT
baryogenesis, in a copious
production of topological defects and in  a
secondary stage of  inflation after  reheating.

\end{quotation}

\newpage

\section{Inflation with $\Omega \not = 1$}
 One of the most robust predictions of inflationary cosmology is that the
universe after inflation becomes extremely flat, which corresponds to $\Omega =
1$. Here $\Omega = {\rho\over \rho_c}$,\,  $\rho_c$ being the energy density of
a
flat universe.  There were many good reasons to believe that this prediction
was quite generic.  The only way to avoid this conclusion is to assume that the
universe inflated only by about $e^{60}$ times. Exact value of the number of
e-folds $N$ depends on details of the theory and may somewhat differ from
60. It is important, however, that in any particular theory  inflation by
extra 2 or 3 e-foldings would make  the universe with $\Omega = 0.5$ or with
$\Omega = 1.5$ almost exactly flat. Meanwhile, the typical number of
e-foldings   in chaotic inflation scenario in the theory ${m^2\over 2}
\phi^2$ is not 60 but rather $10^{12}$.

One can construct models where
inflation leads to expansion of the universe by the factor $e^{60}$. However,
in most of such models small number of e-foldings simultaneously implies that
  density perturbations
are extremely large.   It may be possible to overcome   this obstacle by a
specific choice of the
effective potential. However, this would be only a partial solution. If
the universe
does not inflate long enough to become flat, then by the same token it
does not inflate long enough to become homogeneous and isotropic.
Thus,   the main reason why it is difficult to construct inflationary models
with $\Omega \not = 1$ is not the issue of fine tuning of the parameters of the
models, which is necessary to obtain the universe inflating exactly $e^{60}$
times, but the problem of obtaining a homogeneous universe after inflation.

Fortunately, it is possible to solve this problem, both for a closed universe
(Linde 1992) and for an open one (Coleman and De Luccia, 1980, Gott 1982,
Sasaki et al, 1993). The
main idea is to use the well known fact that the region of space created in the
process of a quantum tunneling tends to have a spherically symmetric shape,
and homogeneous interior, if the tunneling process is suppressed strongly
enough. Then such bubbles of a new phase  tend  to evolve (expand) in a
spherically symmetric
fashion. Thus, if one
could associate the whole visible part of the universe with an interior of one
such region, one would solve the homogeneity problem, and then all other
problems
will be solved by the subsequent relatively short stage of inflation.

For a closed universe the realization of this program is relatively
straightforward (Linde, 1992, 1995). One should consider the process of quantum
creation of  a
closed inflationary universe from ``nothing.''  If the probability of such a
process is exponentially suppressed (and this is indeed the case if inflation
is possible only at the energy density much smaller than the Planck density
(Linde, 1984, Vilenkin, 1984), then the universe created that way will be
rather
homogeneous from the very beginning.

The situation with an open universe is much more complicated. Indeed, an open
universe is infinite, and it may seem impossible to create an infinite universe
by a tunneling process. Fortunately, this is not the case: any bubble formed in
the process of the false vacuum decay looks from inside like an infinite open
universe (Coleman and De Luccia, 1980, Gott 1982, Sasaki {\it et al}, 1993).
 If this universe continues inflating
inside the bubble (Gott 1982, Bucher {\it et al}, 1995) then we obtain an open
inflationary
universe.

There is an extensive investigation of the one-bubble open universe scenario,  
and many important results have been obtained, see e.g.  

(Tanaka and Sasaki, 1994, Sasaki {\it et al}, 1995,  Yamamoto {\it et al},
1995, Bucher {\it et al}, 1995, Bucher and Turok, 1995, Hamazaki {\it et al},
1995). However, for a long time it was not quite clear whether it is
possible to realize this scenario in a natural way.   It would be very nice to
to obtain an open universe
in a theory of just one scalar  field, but in practice it is
rather difficult to obtain a satisfactory model of this type. Typically one is
forced either to introduce very complicated effective potentials, or consider
theories with nonminimal kinetic terms for the inflaton field.
This makes the models   fine-tuned and complicated. It  is very good to know
that  the
models of such type in principle can be constructed, but  it is also very
tempting to find a
more natural realization of the   inflationary universe scenario which would
give
inflation with $\Omega < 1$.

Fortunately, this goal
can be easily achieved if one considers models of two
scalar fields (Linde, 1995, Linde and Mezhlumian, 1995, Garc\'{\i}a--Bellido,
1995). One of these fields may be the standard inflaton
field
$\phi$ with a relatively small mass, another may be, e.g., the scalar field
responsible for the symmetry breaking in GUTs. The presence of two scalar
fields allows one to obtain the required bending of the inflaton potential by
simply changing the definition of the inflaton field in the process of
inflation. At the first stage the role of the inflaton is played by a heavy
field with a steep barrier in its potential, while on the second stage the
role of the inflaton is played by a light field, rolling in a flat direction
``orthogonal'' to the direction of quantum tunneling. This change of the
direction of evolution in the space of scalar fields removes the naturalness
constraints for the form of the potential, which are present in the case of one
field.

Inflationary models of this type
are quite simple, yet they have many interesting features. In these models
the universe consists of infinitely many expanding bubbles immersed into
exponentially expanding false vacuum state. Each of these bubbles inside looks
like an open universe, but the values of $\Omega$ in these universes may take
any value from $1$ to $0$.
In some of these models the situation is even more complicated: Interior of
each bubble looks like an infinite  universe with an effective value of
$\Omega$
slowly decreasing to $\Omega = 0$ at an exponentially  large distance from the
center  of the bubble. We will call such universes quasiopen. Thus, rather
unexpectedly, we are obtaining a large  variety of  interesting and previously
unexplored possibilities.

Here we will describe an extremely  simple model of two
scalar fields, where the universe after inflation becomes open (or quasiopen,
see below) in a very natural way  (Linde, 1995, Linde and Mezhlumian, 1995).

Consider a model of
two noninteracting scalar fields, $\phi$ and $\sigma$, with the effective
potential
\begin{equation}\label{3}
V(\phi, \sigma) = {m^2\over 2}\phi^2 + V(\sigma) \ .
\end{equation}
Here $\phi$ is a weakly interacting inflaton field, and $\sigma$, for example,
can be the field responsible for the symmetry breaking in GUTs. We will assume
that $V(\sigma)$ has a local minimum at $\sigma = 0$, and a global minimum at
$\sigma_0 \not = 0$, just as in the old inflationary
theory. For definiteness, we will assume that this potential is given by
${M^2\over 2} \sigma^2 -
{\alpha M } \sigma^3 + {\lambda\over 4}\sigma^4 + V(0)$, with $V(0) \sim
{M^4\over 4 \lambda}$, but it is not essential;
no fine tuning of the shape of this potential will be required.

Note that so far we did not make any unreasonable complications to the standard
chaotic inflation scenario; at large $\phi$ inflation is driven
by the field $\phi$, and the GUT potential is necessary in the theory anyway.
In order to obtain density perturbations of the necessary amplitude the mass
$m$ of the scalar field $\phi$ should be of the order of $10^{-6} M_{\rm P}
\sim
10^{13}$ GeV (Linde, 1990).

Inflation begins at $V(\phi, \sigma) \sim M_{\rm P}^4$. At this stage
fluctuations of
both fields are very strong, and the universe enters the stage of
self-reproduction, which finishes for the field $\phi$ only when it becomes
smaller than $M_{\rm P} \sqrt{M_{\rm P}\over m}$ and the energy density drops
down to $m
M_{\rm P}^3  \sim 10^{-6} M_{\rm P}^4$ (Linde, 1990). Quantum fluctuations of
the field
$\sigma$ in some parts of the universe put it directly to the absolute minimum
of
$V(\sigma)$, but in some other parts the scalar field $\sigma$ appears in the
local minimum of $V(\sigma)$ at $\sigma  = 0$. We will follow evolution of such
domains. Since the energy density in such
domains will be greater, their volume will  grow  with a greater speed, and
therefore they will be especially important for us.

One may worry that all
domains with $\sigma = 0$
will tunnel to the minimum of $V(\sigma)$ at the stage when the field $\phi$
was very large and quantum fluctuations of the both fields were large too.
This may happen if the Hubble constant induced by the scalar field $\phi$ is
much greater than the curvature of the potential $V(\sigma)$:
\begin{equation}\label{s1}
{m\phi\over M_{\rm P}} {\
\lower-1.2pt\vbox{\hbox{\rlap{$>$}\lower5pt\vbox{\hbox{$\sim$}}}}\ } M \ .
\end{equation}

This decay can be easily suppressed if one introduces a
small interaction $g^2\phi^2\sigma^2$ between these two fields, which
stabilizes the state with $\sigma = 0$ at large $\phi$. Another possibility is
to add a
nonminimal interaction with gravity of the form $-{\xi\over 2} R\phi^2$, which
makes inflation impossible for $\phi > {M_{\rm P}\over 8\phi\xi}$. In this case
the
condition (f{s1}) will never be satisfied.  However, there is a much simpler
answer to this worry. If the effective potential of the field $\phi$ is so
large that the field $\sigma$ can easily jump to the true minimum of
$V(\sigma)$,
then the universe becomes divided into infinitely many domains with all
possible values of $\sigma$ distributed in the following way
(Linde, 1990):
\begin{equation}\label{s2}
{P(\sigma= 0)\over P(\sigma = \sigma_0)} \sim \exp\left({3M^4_{\rm P}\over 8
V(\phi,0)} - {3M^4_{\rm P}\over 8V(\phi,\sigma)}\right) = \exp\left({3M^4_{\rm
P}\over 4(m^2\phi^2 + 2V(0))} - {3M^4_{\rm P}\over 4 m^2\phi^2}\right)\ .
\end{equation}
One can easily check that at the moment when the field $\phi$ decreases to ${M
M_{\rm P}\over m}$ and  the condition (f{s1}) becomes violated, we will  have
\begin{equation}\label{s3}
{P(0)\over P(\sigma_0)}  \sim \exp\left(-{C\over \lambda}\right) \ ,
\end{equation}
where $C$ is some constant, $C = O(1)$. After this moment the probability of
the false vacuum decay typically becomes much smaller. Thus the fraction of
space which survives in the false vacuum state $\sigma = 0$ until this time
typically is very small, but finite (and calculable). It is important,   that
these rare domains with $\sigma = 0$ eventually will dominate the volume of the
universe since if the probability of the false vacuum decay is small enough,
the volume of the domains in the false vacuum will continue growing
exponentially without end.

The main idea of our scenario can be explained as follows. Because the fields
$\sigma$ and
$\phi$ do not interact with each other, and the dependence of the probability
of tunneling on the vacuum energy at the GUT scale is negligibly small
(Coleman and De Luccia, 1980), tunneling to the minimum of $V(\sigma)$ may
occur with approximately
equal
probability at all sufficiently small values of the field $\phi$ (see, however,
below). The
parameters of the bubbles of the field $\sigma$ are determined by the mass
scale $M$ corresponding to the effective potential $V(\sigma)$. This mass scale
in
our model is much greater than $m$. Thus the duration of tunneling in the
Euclidean ``time'' is much smaller than $m^{-1}$. Therefore the field $\phi$
practically does not change its value during the tunneling.  If
the probability of decay at a given $\phi$ is small enough, then it does not
destroy the whole vacuum state $\sigma = 0$; the bubbles of the new
phase are produced all the way when    the field $\phi$ rolls down to $\phi =
0$. In this process  the universe  becomes filled with
(nonoverlapping) bubbles immersed in the false vacuum state with $\sigma = 0$.
Interior of each of these bubbles   represents an open universe. However, these
bubbles   contain  {\it  different} values of the field $\phi$, depending on
the
value of this field at the  moment when the bubble formation occurred. If the
field $\phi$ inside a bubble is smaller than $3 M_{\rm P}$, then the universe
inside
this bubble will have a vanishingly small $\Omega$, at the age $10^{10}$ years
after the end of inflation it will be practically empty, and life of our type
would not exist there.  If the field $\phi$ is much greater than $3 M_{\rm P}$,
the
universe inside the bubble will be almost exactly flat, $\Omega = 1$, as in the
simplest version of the chaotic inflation scenario. It is important, however,
that  {\it  in  an eternally existing self-reproducing universe there will be
infinitely many universes containing any particular value of $\Omega$, from
$\Omega = 0$ to $\Omega = 1$}, and one does not need any fine tuning of the
effective potential to obtain a universe with, say,  $0.2 <\Omega < 0.3$

Of course, one can argue that we did not solve the problem of fine tuning, we
just transformed it into the fact that only a very small percentage of all
universes will have  $0.2 <\Omega < 0.3$. However, first of all, we
achieved our goal in a very simple theory, which does not require any
artificial potential bending and nonminimal kinetic terms. Then, there may be
some reasons why it is preferable for us to live in a universe with a small
(but not vanishingly small) $\Omega$.

The simplest way to approach this problem is to find  how the probability
for the bubble production depends on $\phi$. As we already pointed out, for
small $\phi$ this dependence is not very strong. On the other hand, at large
$\phi$ the probability rapidly grows and  becomes quite large at $\phi > {M
M_{\rm P}\over m}$. This may suggest that the bubble production typically
occurs at
$\phi > {M M_{\rm P}\over m}$, and then for ${M\over  m} \gg 3$ we typically
obtain
flat universes, $\Omega = 1$. This is another manifestation of the problem of
premature decay of the state $\sigma = 0$ which we discussed above. Moreover,
even if the probability to produce the universes with different $\phi$ were
entirely $\phi$-independent, one could argue that the main volume of the
habitable parts of the universe is contained in the bubbles with $\Omega = 1$,
since  the interior of each such bubble  inflated longer. Indeed, the total
volume of each bubble created in a state with  the field $\phi$ during
inflation in our model grows by the factor of $\exp{6\pi\phi^2\over M_{\rm
P}^2}$ (Linde, 1990). It seems clear that the bubbles with greater $\phi$ will
give the largest contribution to the total volume of the universe after
inflation. This would be the simplest argument in favor of the standard
prediction $\Omega = 1$ even in our   class of models.

However, there exist
several ways of resolving this problem: involving coupling $g^2\phi^2\sigma^2$,
which stabilizes the state $\sigma = 0$ at large $\phi$, or adding nonminimal
interaction with gravity of the form $-{\xi\over 2} R\phi^2$. In either way one
can
easily suppress production of the universes with   $\Omega = 1$. Then the
maximum of probability will correspond to some value $\Omega < 1$, which can be
made equal to any given number from $1$ to $0$ by changing the parameters $g^2$
and $\xi$.

For example, let us   add  to the Lagrangian the term $-{\xi\over 2} R\phi^2$.
This term makes inflation impossible for $\phi > \phi_c = {M_{\rm P}\over
\sqrt{8\pi\xi}}$. If initial value of the field $\phi$ is much smaller than
$\phi_c$, the size of the universe during inflation grows $\exp{2\pi\phi^2\over
M_{\rm P}^2}$ times, and the volume grows $\exp{6\pi\phi^2\over M_{\rm P}^2}$
times, as in the theory ${m^2\over 2} \phi^2$ with $\xi = 0$. For initial
$\phi$ approaching $\phi_c$ these expressions somewhat change, but in order to
get a very rough estimate of the increase of the size of the universe in this
model (which is sufficient to get an illustration of our main idea) one can
still use the old expression $\exp{2\pi\phi^2\over M_{\rm P}^2}$. This
expression reaches its maximum near $\phi = \phi_c$, at which point the
effective gravitational constant becomes infinitely large and inflationary
regime ceases to exist (Futamase, 1989, Garc\'{\i}a--Bellido and Linde, 1995).
Thus, one may argue that in this case the main part of the volume of the
universe will appear from the bubbles with initial value of the field $\phi$
close to $\phi_c$. For $\xi \ll 4.4\times 10^{-3}$ one has $\phi_c \gg 3 M_{\rm
P}$. In this case  one would have typical universes expanding much more than
$e^{60}$ times, and therefore $\Omega \approx 1$. For    $\xi \gg 4.4\times
10^{-3}$ one has $\phi_c \ll 3 M_{\rm P}$, and therefore  one would have
$\Omega \ll 1$ in all inflationary bubbles. It is clear that by choosing
particular values of the constant  $\xi$ in the range of  $\xi \sim 4.4\times
10^{-3}$ one can obtain the distribution of the universes with the maximum of
the distribution concentrated near any desirable value of $\Omega < 1$.
Note that the position of the peak of the distribution is very sensitive to the
value of $\xi$: to have the peak concentrated in the region $0.2 < \Omega <
0.3$ one would have to fix $\xi$ (i.e. $\phi_c$) with an accuracy of few
percent. Thus, in this approach to the calculation of probabilities to live in
a universe with a given value of $\Omega$ we still have the problem of fine
tuning.

However, calculation of probabilities in the context of the theory of a
self-reproducing universe is a very ambiguous process, and it is even not quite
clear that this process makes any sense at all. For example, we may
formulate the problem in a different way. Consider a domain of the false vacuum
with $\sigma = 0$ and $\phi = \phi_1$. After some evolution it
produces one or many bubbles with $\sigma = \sigma_0$ and the field $\phi$
which after some time becomes equal to $\phi_2$. One may argue that the most
efficient way this process may go is the way which in the end produces the
greater volume. Indeed, for the inhabitants of a bubble it does not matter how
much time did it take for this process to occur. The total number of
observers produced by this process will depend on the total volume of the
universe at the hypersurface of a given density, i.e. on the hypersurface of a
given $\phi$. If the domain instantaneously  tunnels to the state $\sigma_0$
and $\phi_1$, and then the field $\phi$ in this domain slowly rolls from
$\phi_1$ to $\phi_2$, then the volume of this domain grows $\exp
\Bigl({2\pi\over M_{\rm P}^2} (\phi_1^2 -\phi_2^2)\Bigr)$ times (Linde, 1990).
Meanwhile, if the tunneling takes a long time, then the field $\phi$ rolls down
extremely slowly being in the false vacuum state with $\sigma = 0$. In this
state the universe expands much faster than in the state with $\sigma =
\sigma_0$. Since it expands much faster, and it takes the field much longer to
roll from $\phi_1$ to $\phi_2$, the trajectories of this kind bring us much
greater volume. This may serve as an argument that most of the volume is
produced by the bubbles created at a very small $\phi$, which leads to the
universes with very small $\Omega$.

One may use another set of considerations, studying all trajectories beginning
at $\phi_1, t_1$ and ending at $\phi_2, t_2$. This will bring us  another
answer, or, to be more precise, another set of answers, which will depend on
the choice of the time parametrization (Linde {\it et al}, 1994).  Still
another answer will be obtained by the method
  recently proposed by Vilenkin, who suggested to introduce a particular
cutoff procedure which partially eliminates dependence of the final
answer on the time parametrization (Vilenkin, 1995, Winitzki and Vilenkin,
1995)). However, there exists a wide class of cutoff procedures which have
similar properties, but give exponentially different results (Linde and
Mezhlumian, 1995a)

There is a very deep  reason why the calculation of the probability to obtain a
universe with a given $\Omega$ is so ambiguous. We have discussed this reason
in Sect. 3.1 in general terms; let us see how the situation looks in
application to the open universe scenario. For those who  lives inside
a bubble there is be no way to say at which stage (at which
time from the point of view of an external observer) this bubble was produced.
Therefore one should compare  {\it  all} of these bubbles produced at all
possible times.  The self-reproducing universe should exist for indefinitely
long time, and therefore   it should contain  infinitely many bubbles with all
possible values of $\Omega$. Comparing infinities is a very ambiguous task,
which gives results depending on the procedure of comparison. For example, one
can consider an infinitely large box of apples and an infinitely large box of
oranges. One may pick up one apple and one orange, then one apple and one
orange, over and over again, and conclude that there is an equal number of
apples and oranges. However, one may also pick up one apple and
two oranges, and then one apple and two oranges again, and conclude that there
is twice as many oranges as apples. The same situation happens when one tries
to compare the number of  bubbles with different values of $\Omega$. If we
would know how to
solve  the problem of measure in quantum cosmology, perhaps we would be able to
obtain
something similar to an open universe in the trivial $\lambda\phi^4$ theory
without any first order phase transitions (Linde {\it et al} 1995,  1995a).  In
the meantime, it is already encouraging that in our scenario
there are infinitely many inflationary universes with all possible value of
$\Omega < 1$. We can hardly live in the empty bubbles with $\Omega = 0$. As for
the choice between the bubbles with different nonvanishing values of $\Omega <
1$,  it is quite possible that eventually we will find out an unambiguous way
of predicting the most probable value of $\Omega$, and we are going to continue
our work in this direction. However,  as we already discussed in the previous
section, it might also happen that this question
is as meaningless as the question whether it is more probable to be born as a
Chinese rather than as an Italian. It is quite conceivable that the only way to
find out in which of the bubbles do we   live   is to make observations.

Some words of caution are in order here. The bubbles produced in our simple
model
are not  {\it  exactly} open universes. Indeed, in the one-field models  the
time of reheating (and the temperature of the
universe after the reheating) was exactly synchronized with the value of the
scalar
field inside the bubble. In our case the situation is very similar, but not
exactly. Suppose that the Hubble constant induced by $V(0)$ is much
greater than the Hubble constant related to the energy density of the scalar
field $\phi$. Then the speed of rolling of the scalar field $\phi$ sharply
increases inside the bubble. Thus, in our case the field $\sigma$ synchronizes
the motion of the field $\phi$, and then the hypersurface of a constant field
$\phi$ determines the hypersurface of a constant temperature. In the models
where the rolling of the field $\phi$ can occur only inside the bubble (we will
discuss such a model shortly)  the  synchronization is precise, and everything
goes as in the one-field models. However, in our simple
model the scalar field $\phi$ moves down outside the bubble as well, even
though it does it very slowly. Thus,  synchronization of   motion of the
fields $\sigma$ and $\phi$  is not precise; hypersurface of a constant $\sigma$
ceases to be a hypersurface of a constant density. For example, suppose that
the field $\phi$ has taken some value $\phi_0$ near the bubble wall when the
bubble was just formed. Then the bubble expands, and during this time the field
$\phi$ outside the wall  decreases, as $\exp \Bigl(-{m^2t\over 3 H_1}\Bigr)$,
where $H_1 \approx  H(\phi = \sigma = 0)$ is the Hubble constant at the first
stage of inflation, $H_1 \approx \sqrt{8\pi V(0)\over 3 M_{\rm P}^2}$. At the
moment
when the bubble expands $e^{60}$ times, the field $\phi$ in the region just
reached by  the bubble wall decreases to  $\phi_o\exp \Bigl(-{20 m^2\over
H^2_1}\Bigr)$ from its original value $\phi_0$. the universe inside the bubble
is a homogeneous open universe only if this change is negligibly small. This
may not be a real problem. Indeed,  let us assume that $V(0) ={\tilde M}^4$,
where ${\tilde M} =
10^{17}$ GeV. (Typically the energy density scale $\tilde M$ is related to the
particle mass as follows: ${\tilde M} \sim \lambda^{-1/4} M$.) In this case
$H_1 = 1.7 \times 10^{15}$ GeV, and for $m =
10^{13}$ GeV one obtains ${20 m^2\over   H_1^2} \sim 10^{-4}$. In such a case
a typical degree of distortion of the picture of a homogeneous open universe is
very small.

Still this issue requires careful investigation. When the bubble wall continues
expanding even further, the scalar field outside of it eventually drops down to
zero. Then there will be no new matter created near the wall.  Instead of
infinitely large homogeneous open universes we are obtaining   spherically
symmetric islands of a size much greater than the size of the observable part
of our universe. We do not know whether this unusual picture is an advantage or
a
disadvantage of our model. Is it possible to consider different parts of the
same
exponentially large island as domains of different ``effective'' $\Omega$? Can
we attribute some part of the dipole anisotropy of the microwave background
radiation to the possibility that we live somewhere outside of the center of
such island? In any
case, as we already mentioned, in the limit $m^2 \ll H_1^2$
we do not expect that the small deviations of the geometry of space inside the
bubble from the geometry of an open universe can do much harm to our model.

Our model admits many generalizations, and details of the scenario which we
just discussed depend on the values of parameters. Let us forget for a moment
about all complicated processes which occur  when the field $\phi$ is rolling
down to $\phi = 0$, since this part of the picture depends on the validity of
our ideas about initial conditions. For example, there may be no
self-reproduction of inflationary domains with large $\phi$ if one considers an
effective
potential of the field $\phi$ which is very curved at large $\phi$. However,
there will be self-reproduction of the universe in a state
$\phi = \sigma = 0$, as in the old inflation scenario. Then the main portion of
the volume of the universe will be determined by the processes which occur when
   the fields    $\phi$  and $\sigma$    stay  at the local minimum of the
effective potential, $\phi = \sigma = 0$.   For definiteness we will assume
here that $V(0) = {\tilde M}^4$, where ${\tilde M}$ is   the   stringy scale,
${\tilde M} \sim 10^{17} -
10^{18}$ GeV. Then the Hubble constant $H_1 = \sqrt{8\pi V(0)\over 3M^2_{\rm
P}} \sim
\sqrt{8\pi \over 3} {{\tilde M}^2\over M_{\rm P}}$ created by the energy
density
$V(0)$ is
much greater than $m \sim 10^{13}$ GeV. In such a case the scalar field $\phi$
will not stay exactly at $\phi = 0$. It will  be  relatively homogeneous on the
horizon scale $H_1^{-1}$, but otherwise it will  be chaotically distributed
with
the dispersion $\langle\phi^2\rangle = {3H^4\over 8\pi^2m^2}$ (Linde, 1990).
This means that the field $\phi$ inside each of the bubbles produced by the
decay of the false vacuum can take any value $\phi$ with the probability
\begin{equation}\label{4}
P \sim \exp\left(-{\phi^2\over 2 \langle\phi^2\rangle}\right) \sim
\exp\left(-{3m^2 \phi^2M_{\rm P}^4\over 16 {\tilde M}^8}\right) \ .
\end{equation}
One can check that for ${\tilde M} \sim 4.3\times10^{17}$ GeV the typical value
of the
field $\phi$ inside the bubbles will be $\sim 3\times 10^{19}$ GeV. Thus, for
${\tilde M} > 4.3\times10^{17}$ GeV most of the universes produced during the
vacuum
decay will be flat, for ${\tilde M} < 4.3\times10^{17}$ GeV most of them will
be open.
It is interesting that in this version of our model the percentage of open
universes is determined by the stringy scale (or by the GUT scale). However,
since the process of bubble production in this scenario   goes without   end,
the total  number of universes with any particular value of  $\Omega < 1$ will
be infinitely large   for any value of ${\tilde M}$.   Thus this   model shows
us is the
simplest way to resurrect some of the ideas of the old inflationary theory with
the help of chaotic inflation, and simultaneously to obtain  inflationary
universe with $\Omega < 1$.

Note that this version of our model will not suffer for the problem of
incomplete synchronization. Indeed, the average value of the field $\phi$ in
the false vacuum outside the bubble will remain constant until the bubble
triggers its decrease.
However, this model, just as its previous version, may suffer from another
problem. The Hubble constant $H_1$ before the tunneling in this model was much
greater
than the Hubble constant $H_2$ at the beginning of the second stage of
inflation. Therefore the fluctuations of the
scalar field before the tunneling were very large, $\delta \phi \sim {H_1\over
2 \pi}$, much greater than the
fluctuations generated after the tunneling,  $\delta \phi \sim {H_2\over 2
\pi}$. This may lead to very large
density perturbations on the scale comparable to the size of the bubble. For
the models with $\Omega = 1$ this effect would not cause any problems since
such
perturbations would be far away over the present particle horizon, but for
small $\Omega$  this
may lead to unacceptable anisotropy of the microwave background radiation.

Fortunately, this may not be a real difficulty. A possible solution is very
similar to the bubble symmetrization described in the previous section.

Indeed, let us consider more carefully how the long wave perturbations produced
outside the bubble may penetrate into it. At the moment when the bubble is
formed, it has a   size    smaller than $H_1^{-1}$
(Coleman and De Luccia, 1980). Then the bubble walls begin moving with the
speed gradually
approaching the speed of light. At this stage the comoving size of the bubble
(from the point of view of the original coordinate system in the false vacuum)
grows like
\begin{equation}\label{n1}
r(t) = \int_{0}^{t}{dt e^{-H_1 t}} = H_1^{-1} (1 - e^{-H_1 t}) \ .
\end{equation}
During this time the fluctuations of the scalar field $\phi$ of the amplitude
${H_1\over 2\pi}$ and of the wavelength $H_1^{-1}$, which previously were
outside the bubble, gradually become covered by it. When these perturbations
are outside the bubble, inflation with the Hubble constant $H_1$ prevents them
from oscillating and moving. However, once these perturbations penetrate inside
the bubble, their amplitude becomes decreasing (Mukhanov and Zelnikov, 1991).
Indeed, since the
wavelength of the perturbations is $\sim H_1^{-1} \ll H_2^{-1} \ll m^{-1}$,
these
perturbations move inside the bubbles as relativistic particles, their
wavelength grow  as $a(t)$, and their amplitude decreases just like an
amplitude of electromagnetic field, $\delta\phi \sim a^{-1}(t)$, where $a$ is
the scale factor of the universe inside a bubble (Mukhanov and Zelnikov, 1991).
This process
continues until the wavelength of each perturbation reaches $H_2^{-1}$ (already
at the second stage of inflation). During this time the wavelength grows
${H_1\over H_2}$ times, and the amplitude decreases ${H_2\over H_1}$ times, to
become the standard amplitude of perturbations produced at the second stage of
inflation: $  {H_2\over H_1}\, {H_1\over 2\pi} = {H_2\over 2\pi}$.

In fact, one may  argue that this computation was too naive, and that these
perturbations should be neglected altogether. Typically we treat long wave
perturbations in inflationary universe like classical wave for the reason that
the waves with the wavelength much greater than the horizon can be interpreted
as states with extremely large occupation numbers (Linde, 1990). However, when
the new  born perturbations (i.e. fluctuations which did not acquire an
exponentially large wavelength yet) enter  the bubble (i.e. under the horizon),
they effectively return to the realm of quantum fluctuations again. Then one
may argue that one should simply forget about the waves with the wavelengths
small enough to fit into the bubble, and consider perturbations created at the
second stage of inflation not as a result of stretching of these waves, but as
a new process of creation of perturbations of an amplitude ${H_2\over 2\pi}$.

One may worry   that perturbations which had wavelengths somewhat greater than
$H_1^{-1}$ at the moment of the bubble formation  cannot completely penetrate
into the bubble. If, for example, the field $\phi$ differs from some constant
by $+{H_1\over 2\pi}$ at the distance $H_1^{-1}$ to the left   of the bubble at
the moment of its formation, and by  $-{H_1\over 2\pi}$ at the distance
$H_1^{-1}$ to the  right of the bubble, then this difference remains frozen
independently of all processes inside the bubble. This may suggest that there
is some unavoidable asymmetry of the distribution of the field inside the
bubble. However, the field inside the bubble will not be distributed like a
straight line slowly rising from  $-{H_1\over 2\pi}$ to  $+{H_1\over 2\pi}$.
Inside
the bubble the field will be almost homogeneous; the inhomogeneity $\delta \phi
\sim -{H_1\over 2\pi}$ will be concentrated only in a small vicinity near the
bubble wall.

Finally we should verify that this scenario  leads to bubbles which are
symmetric enough. Fortunately, here we do not have any
problems. One can easily check that for our model with $m \sim 10^{13}$ GeV and
$\tilde M \sim \lambda^{-1/4} M > 10^{17} GeV$ perturbations of metric induced
by the wall perturbations are small    even for not very small values of the
coupling constant $\lambda$ (Linde and Mezhlumian, 1995, Garc\'{\i}a--Bellido,
1995).

The arguments presented above should be confirmed by a more detailed
investigation of the vacuum structure inside the expanding bubble in our
scenario. If, as we hope,  the result of the investigation will be positive, we
will have an
extremely simple model of an open inflationary universe. In the meantime, it
would be nice to have a model where we do not have any problems at all with
synchronization and
with  large fluctuations on the scalar field in the false vacuum.

The simplest model of this kind is a version of the hybrid
inflation scenario (Linde, 1991, 1994), which is a slight generalization (and a
simplification) of our previous model
(f{3}):
\begin{equation}\label{4a}
V(\phi,\sigma) = {g^2\over 2}\phi^2\sigma^2 + V(\sigma) \ .
\end{equation}
We eliminated the massive term of the field $\phi$ and added explicitly the
interaction ${g^2\over 2}\phi^2\sigma^2$, which, as we have mentioned already,
can be useful (though not necessary)  for stabilization of the state $\sigma =
0$ at large $\phi$. Note
that in this model the line $\sigma = 0$ is a flat direction in the
($\phi,\sigma$) plane. At large $\phi$ the only minimum of the effective
potential with respect to $\sigma$ is at the line $\sigma = 0$.  To give a
particular example, one can take $V(\sigma) = {M^2\over 2} \sigma^2 -{\alpha M
} \sigma^3 + {\lambda\over 4}\sigma^4 +V_0$. Here $V_0$ is a constant which is
added to ensure that $V(\phi,\sigma) = 0$ at the absolute minimum of
$V(\phi,\sigma)$.  In this case the minimum of the potential $V(\phi,\sigma)$
at $\sigma \not = 0$ is deeper than the minimum at $\sigma = 0$ only for $\phi
< \phi_c$, where $\phi_c = {M\over g}\sqrt{{2\alpha^2\over  \lambda} -1}$. This
minimum for $\phi = \phi_c$ appears at $\sigma = \sigma_c = {2\alpha M\over
\lambda}$.

The bubble formation becomes possible only for $\phi < \phi_c$. After the
tunneling the field $\phi$ acquires an effective mass $m = g\sigma$ and begins
to move towards $\phi = 0$, which provides the mechanism for the second stage
of inflation inside the bubble. In this scenario evolution of the scalar field
$\phi$ is exactly synchronized with the evolution of the field $\sigma$, and
the universe inside the bubble appears to be open.

Effective mass of the   field $\phi$ at the minimum of $V(\phi,\sigma)$ with
$\phi = \phi_c$, $\sigma = \sigma_c = {2\alpha M\over  \lambda}$ is   $m =
g\sigma_c = {2g\alpha M\over  \lambda}$. With a decrease of the field $\phi$
its effective mass at the minimum of $V(\phi,\sigma)$ will grow, but not
significantly. For simplicity, we will consider the case $\lambda = \alpha^2$.
 In this case it can be shown that $V(0) = 2.77\, {M^4\over \lambda}$, and the
Hubble constant before the phase transition is given by $4.8\, {M^2\over \sqrt
\lambda M_{\rm P}}$.   The effective mass $m$ after the phase transition is
equal to ${2gM\over \sqrt
\lambda}$ at $\phi = \phi_c$, and then it grows by only $25\%$ when the field
$\phi$ changes all the way down from  $\phi_c$ to $\phi = 0$.

The bubble
formation becomes possible only for $\phi < \phi_c$. If it happens in the
interval $4M_{\rm P} > \phi > 3 M_{\rm P}$, we obtain a flat universe. If it
happens at $\phi < 3M_{\rm P}$, we obtain an open universe. Depending on the
initial value of the field $\phi$, we can obtain all possible values of
$\Omega$, from $\Omega = 1$ to $\Omega = 0$. The value of the Hubble constant
at the minimum with $\sigma \not = 0$ at $\phi = 3M_{\rm P}$ in our model does
not differ much from the value of the Hubble constant before the bubble
formation. Therefore we do not expect any specific problems with the large
scale density perturbations in this model.
 Note also that the probability of tunneling at large $\phi$ is very small
since the depth of the minimum at $\phi \sim \phi_c$, $\sigma \sim \sigma_c$
does not differ much from the depth of the minimum at $\sigma = 0$, and there
is no tunneling at all for $\phi > \phi_c$. Therefore
the number of flat universes produced by this mechanism will be strongly
suppressed as compared with the number of open universes, the degree of this
suppression being very sensitive to the value of $\phi_c$. Meanwhile, life of
our type is impossible in empty universes with $\Omega \ll 1$. This may provide
us with a tentative explanation of the small value of $\Omega$ in the context
of our model.

Another model  of inflation with $\Omega < 1$ is the based on a certain
modification of the ``natural inflation'' scenario (Freese {\it et al}, 1990).
The main idea is to take the effective potential of the ``natural inflation''
model, which looks like a tilted Mexican hat,  and make a deep hole in its
center at $\phi = 0$ (Linde and Mezhlumian, 1995). In the beginning inflation
occurs near $\phi = 0$, but then the bubbles with $\phi \not = 0$ appear.
Depending on the phase of the complex scalar field $\phi$ inside the bubble,
the next stage of inflation, which occurs just as in the old version of the
``natural inflation'' scenario, leads to formation of the universes with all
possible values of $\Omega$.
Thus, there exist several simple inflationary models which lead to the picture
of the universe consisting of many bubbles with different values of $\Omega$.
Therefore instead of insisting that inflation leads to $\Omega = 1$ or
estimating the probability to live in a bubble with a given value of $\Omega$
we should ask astronomers to measure it.

\section{Nonthermal Phase Transitions after Inflation}

 The theory of reheating is one of the most important parts of
inflationary
cosmology. Elementary theory of this process was developed many years
ago by
Dolgov and Linde (1982) and by Abbott {\it et al} (1982).   Some important
steps toward a complete theory have
been
made in (Dolgov and  Kirilova,  1990, Traschen and  Brandenberger,  1990).
However,  the real progress in
understanding of
this process was achieved only recently when the new theory of
reheating was
developed.  According to this theory (Kofman {\it et al}, 1994), reheating
typically consists
of  three different stages.  At the first stage, a classical
 oscillating scalar  field $\phi$ (the  inflaton field) decays into
massive
bosons due to parametric
resonance.  In
many models the resonance is very broad, and the process occurs
extremely
rapidly.  To distinguish this stage of explosive reheating from the
stage of  particle decay and
thermalization, we called it {\it preheating}.
Bosons produced at that stage are far away from thermal equilibrium
and have enormously large occupation numbers. The second stage
is the
decay of previously produced particles. This stage typically can be
described
by the elementary theory developed  by
Dolgov and Linde (1982) and by Abbott {\it et al} (1982). However, these
methods should be
applied
not  to the decay of the original homogeneous inflaton field, but to
the decay
of particles produced at the  stage of preheating. This
 changes many features of the process including the final value of
the
reheating temperature.  The third stage of reheating is
thermalization.

Different aspects of the theory of explosive reheating have been
studied  by many  authors (Shtanov {\it et al},  1995,  Boyanovsky  {\it et
al},  1995, Yoshimura, 1995, Kaiser, 1995, Fujisaki {\it et al},  1995).
In our presentation we will follow  the original approach of Kofman {\it et al}
(1994),
where the theory of reheating was investigated with an account taken  both  of
the expansion of the universe and of the backreaction of created particles. The
results reported here have been obtained by Kofman {\it et al} (1995, 1996).

One should note that
there exist such models where this first stage of reheating is
absent; e.g, there is no parametric resonance in the theories where
the
 field $\phi$ decays into fermions. However, in the theories where
preheating is possible one may expect  many  unusual phenomena.
One of the most
interesting  effects  is the
possibility of specific non-thermal post-inflationary phase
transitions which occur after preheating. As we
will see, these phase transitions in certain cases can be much more
pronounced
that the standard high temperature cosmological phase transitions. They may
lead to copious
production of topological defects and to a
secondary stage of  inflation after  reheating.

 Let us first remember the  theory of  phase transitions in
theories
with spontaneous symmetry breaking (Kirzhnits, 1972,  Kirzhnits. and  Linde,
1972, Weinberg, 1974, Dolan and  Jackiw,   1974, Kirzhnits   and  Linde, 1974,
1976). We will consider first the theory of scalar
fields
$\phi$ and $\chi$ with the effective potential
\begin{equation}\label{p1}
V(\phi,\chi) =   {\lambda\over 4}
(\phi^2-\phi_0^2)^2 + {
1\over2} g^2 \phi^2 \chi^2   \ .
\end{equation}
  Here $\lambda, g \ll 1$  are
coupling constants. $V(\phi,\chi)$ has a minimum at $\phi =
\phi_0$,
$\chi = 0$ and a maximum at $\phi = \chi = 0$ with the curvature
$V_{\phi\phi}
= -m^2= - \lambda\phi_0^2$. This effective potential
acquires corrections due  to quantum (or
thermal)
fluctuations of the scalar fields (Weinberg, 1974, Dolan and  Jackiw,   1974,
Kirzhnits   and  Linde, 1974),
$
\Delta V =  {3\over 2} \lambda  \langle (\delta\phi^2)\rangle \phi^2 +
{g^2\over 2}  \langle (\delta\chi)^2\rangle  \phi^2 + {g^2\over
2}\langle (\delta\phi)^2\rangle
\chi^2 +...,
$
where  the quantum field operator is decomposed as $\hat \phi =
\phi + \delta \phi$ with $\phi \equiv \langle \hat \phi\rangle$,
and we have written only
leading terms depending on $\phi$ and  $\chi \equiv \langle \hat \chi
\rangle$. In
the large temperature limit
$
\langle (\delta\phi)^2\rangle = \langle (\delta\chi)^2\rangle = {T^2\over 12}.
$
The effective mass squared of the field $\phi$
\begin{equation}\label{p4}
m_{\phi ,eff}^2 = -m^2 + 3 \lambda \phi^2 +
3\lambda \langle
(\delta\phi)^2\rangle +  g^2\langle(\delta\chi)^2\rangle
\end{equation}
becomes positive and symmetry  is restored  (i.e. $\phi =0$ becomes
the stable equilibrium point) for $T > T_c$,
where
$T^2_c = {12 m^2\over 3\lambda + g^2} \gg m^2$. At this temperature
the
energy density of the gas of ultrarelativistic particles
is given by
$
\rho = N(T_c) {\pi^2\over 30} T_c^4 = {24\, m^4N(T_c)\pi^2\over 5\,
(3\lambda +
g^2)^2} \
{}.
$
Here $N(T)$ is the effective number of degrees of freedom at large
temperature,
which in realistic situations may vary from $10^2$ to $10^3$.
Note that for   $g^4 < {96 N\pi^2\over 5}
\lambda$ this energy
density   is
greater than the vacuum energy density $V(0) = {m^4\over
4\lambda}$.  Meanwhile, for   $g^4 {\
\lower-1.2pt\vbox{\hbox{\rlap{$>$}\lower5pt\vbox{\hbox{$\sim$}}}}\ }
\lambda$
radiative corrections are important, they lead to creation of a
local
minimum of $V(\phi,\chi)$, and the phase transition occurs from a
strongly
supercooled state (Kirzhnits   and  Linde, 1976). That is why the first models
of
inflation
required supercooling at the moment of the phase transition.

An exception from this rule is given by supersymmetric
theories, where
one may have $g^4 \gg \lambda$ and still have a  potential which is
flat near
the origin due to cancellation of quantum corrections of bosons and
fermions
 (Lyth  and  Stewart, 1995). In such cases thermal energy becomes smaller than
the
vacuum energy
at $T < T_0$, where $T^4_0 = {15 \over 2 N \pi^2}m^2\phi_0^2$. Then
one may
even
have a short stage of inflation which begins at $T \sim T_0$ and ends
at $T =
T_c$. During this time the universe may inflate by the factor
\begin{equation}\label{p5a}
{a_c\over a_0} = {T_0\over T_c} \sim 10^{-1} \Bigr({g^4\over
\lambda}\Bigl)^{1/4}\,
\approx 10^{-1} g\,\sqrt{\phi_0\over  m} .
\end{equation}

In supersymmetric theories with flat
directions $\Phi$
it may be more natural  to consider potentials of the so-called  ``flaton''
fields $\Phi$
without the term  ${\lambda \over 4} \Phi^4$  (Lyth  and  Stewart, 1995):
\begin{equation}\label{p5b}
V(\Phi,\chi) =  - {m^2 \Phi^2\over2}
+{\lambda_1 \Phi^6\over 6 M_p^2} + {m^2 \Phi_0^2\over 3} + {1
\over2} g^2 \Phi^2 \chi^2
 \ ,
\end{equation}
where $\Phi_0 =  \lambda_1^{-1/4} \sqrt{m M_p}$ corresponds to the
minimum of
this potential.  The critical temperature in this theory for
$\lambda_1 \Phi_0^2
\ll g^2 M_p^2$ is   the same   as in the theory (f{p1}) for
$\lambda \ll
g^2$, and expansion of the universe during thermal inflation is
given by $10^{-1}
g\,\sqrt{\Phi_0/m}$, as in eq. (f{p5a}). Existence of this short additional
stage of
``thermal inflation'' is a very interesting effect, which may
be very
useful. In particular, it may provide a solution to the Polonyi field
problem (Lyth  and  Stewart, 1995).

The theory of   cosmological phase transitions is an important part
of the
theory of the evolution of the universe, and during the last twenty
years it was investigated in a very detailed way. However, typically
it
was
assumed that the phase transitions occur in the state of thermal
equilibrium.
Now  we are going to show that similar phase transitions
may occur
even much more efficiently  prior to  thermalization, due to the
anomalously
large expectation values $\langle(\delta\phi)^2\rangle$ and
$\langle(\delta\chi)^2\rangle$
produced during preheating.

We will first consider the model (\ref{p1})   without the scalar
field
$\chi$
and with the amplitude of spontaneous symmetry breaking  $\phi_0 \ll
M_{\rm P}$. In this model inflation occurs during the slow rolling of the
scalar
field $\phi$ from its very large values until it becomes of the order
$M_{\rm P}$.
Then it   oscillates with the  initial amplitude $\phi \sim 10^{-1} M_p$ and
initial frequency $\sim 10^{-1} \sqrt  \lambda
M_{\rm P}$.
Within a few dozen  oscillations it transfers most of its energy $\sim
{\lambda\over 4}10^{-4}
M_{\rm P}^4$ to
its long-wave fluctuations $\langle(\delta \phi)^2\rangle$ in the
regime
of broad parametric resonance   (Kofman {\it et al}, 1994).

The crucial observation   is the following. Suppose that the initial energy
density of oscillations $\sim {\lambda\over 4}10^{-4}
 M_{\rm P}^4$ were instantaneously transferred to thermal energy density $\sim
10^2  T^4$. This would give the reheating
temperature  $T_r \sim 2\times 10^{-2} \lambda^{1/4} M_{\rm P}$, and the scalar
field  fluctuations   $\langle(\delta\phi)^2\rangle
\sim T_r^2/12 \sim 3\times 10^{-5} \sqrt \lambda
M_{\rm P}^2$. Meanwhile particles created during preheating have much smaller
energy  $\sim  10^{-1} \sqrt \lambda
M_{\rm P}$. Therefore if the same energy density ${\lambda\over 4}10^{-4}
 M_{\rm P}^4$ is instantaneously transferred to low-energy particles created
during preheating, their number, and, correspondingly, the amplitude of
fluctuations, will be much greater,    $\langle(\delta\phi)^2\rangle\sim C^2
M_{\rm P}^2$,  where $C^2 \sim 10^{-2} - 10^{-3}$ (Kofman {\it et al}, 1994,
1996).   Thermal fluctuations would lead to symmetry restoration in our
model only
for $\phi_0 {\
\lower-1.2pt\vbox{\hbox{\rlap{$<$}\lower5pt\vbox{\hbox{$\sim$}}}}\
}T_r \sim
10^{-2} \lambda^{1/4} M_{\rm P}  \sim 10^{14}$ GeV  for the realistic value
$\lambda \sim 10^{-13}$ (Linde, 1990).   Meanwhile, according to eq.
(\ref{p4}),   the
nonthermalized
fluctuations $\langle(\delta\phi)^2\rangle\sim M_{\rm P}^2$  may lead to
symmetry
restoration  even if
the symmetry
breaking parameter $\phi_0$  is  as large as $10^{-1} M_{\rm P}$. Thus,  the
nonthermal
symmetry restoration may occur  even in those theories where the symmetry
restoration due to  high temperature effects would be
impossible (Kofman {\it et al}, 1995). (Recently a similar  conclusion was
reached also by Tkachev (1995). However, his investigation was based on an
oversimplified picture of reheating, and  his estimates  differ  considerably
from our results.)

In reality thermalization   takes a  very  long time, which is inversely
proportional to coupling constants. This dilutes the energy density, and the
reheating temperature becomes many orders of magnitude smaller than $10^{14}$
GeV (Linde, 1990). Therefore post-inflationary thermal effects typically cannot
restore symmetry on the GUT scale. Preheating is not instantaneous as well, and
therefore the fluctuations produced at that stage are smaller than $C^2 M_{\rm
P}^2$, but only logarithmically: $\langle(\delta\phi)^2\rangle\sim  C^2 M_{\rm
P}^2\ln^{-2}{1\over \lambda}$ (Kofman {\it et al}, 1995, 1996). For $\lambda
\sim  10^{-13}$ this means than nonthermal perturbations produced at reheating
may restore symmetry on the scale up to $\phi_0 \sim 10^{16}$ GeV.

Later   $\langle(\delta\phi)^2\rangle$ decreases as $a^{-2}(t)$  because of the
expansion of the universe. This leads
to  the  phase transition  with symmetry breaking.
The homogeneous component $\phi (t)$ at the moment of the phase transition
happens to be   significantly
less than $\sqrt{\langle (\delta\phi)^2\rangle}$ due to its decay
in the regime of the narrow parametric resonance after
 preheating  (Kofman {\it et al}, 1994):
$\overline {\phi^2} \propto t^{-7/6} \propto t^{-1/6} \langle (\delta \phi)^2
\rangle$;  bar means averaging over oscillations.

The mechanism of symmetry restoration  described above is very general; in
particular, it explains a surprising behavior of  oscillations of the scalar
field found  numerically  in the
$O(N)$-symmetric model discussed by Boyanovsky {\it et al} (1995). It is
important that during the interval between preheating and the establishing  of
thermal equilibrium the universe could experience a
series of phase transitions which we did not anticipate before. For example,
cosmic strings and textures, which could be an additional source  for the
formation of the large scale
structure of the universe,  should have $\phi_0 \sim 10^{16}$ GeV (Vilenkin and
Shellard, 1994). To produce them by thermal phase transitions in our model one
should have the temperature  after
reheating greater than $10^{16}$ GeV, which is extremely hard to obtain
(Kofman and  Linde,  1987).  Even with an account taken of the stage of
explosive  reheating, the resulting reheating temperature typically remains
many orders of magnitude smaller than $10^{14}$ GeV, since it is mainly
determined by the last stages of reheating where the parametric resonance is
inefficient. Meanwhile, as we see now, fluctuations produced during
the first
stage of reheating are more
than sufficient to restore the symmetry. Then the topological defects can be
produced in a standard way  when the symmetry breaks down again. In other
words, production of
superheavy topological defects  can be easily compatible with inflation.

On the other hand, the topological defect production can be quite dangerous.
For example, the model
(\ref{p1})
of a
one-component real scalar field $\phi$ has a discrete symmetry $\phi
\to -
\phi$.
As a result, after the phase transition induced by fluctuations
$\langle(\delta\phi)^2\rangle$ the universe may become filled with
domain
walls
separating phases $\phi = +\phi_0$ and $\phi = -\phi_0$. This
is expected to lead to a cosmological disaster.

This question requires a more detailed analysis. Even though the
point $\phi = 0$ after preheating becomes a minimum of the effective
potential,
the field $\phi$ continues oscillating around
this minimum. Therefore, at the moment $t_c$ it may
happen to  be either
to the right of the maximum of $V(\phi)$ or to the left of it everywhere in
the universe.
In this case the symmetry breaking will occur
in one preferable direction, and no
domain walls will be produced. A similar mechanism may suppress
production of other topological defects.

However, this would be correct only if the magnitude of fluctuations
$(\delta\phi)^2$ were smaller than the average amplitude of the oscillations
$\overline {\phi^2}$.  In our case fluctuations
$(\delta\phi)^2$   are greater than $\overline {\phi^2}$ (Kofman {\it et al},
1994), and they can have
considerable local deviations from their average value
$\langle(\delta\phi)^2\rangle$. Investigation of this question
shows that  in the theory (\ref{p1})  with $\phi_0 \ll 10^{16}$ GeV
fluctuations destroy the coherent distribution of the oscillating
field $\phi$
and divide the universe into equal number of domains with $\phi = \pm \phi_0$,
which leads to the domain wall problem. This means that in
consistent inflationary models of the type of (\ref{p1}) one  should  have
either $\phi_0 = 0$ (no symmetry breaking), or   $\phi_0 {\
\lower-1.2pt\vbox{\hbox{\rlap{$>$}\lower5pt\vbox{\hbox{$\sim$}}}}\ } 10^{16}$
GeV.

 Now we will consider models where the symmetry breaking occurs for
fields other than the inflaton field $\phi$.
The simplest model  has an effective potential
\begin{equation}\label{p7}
V(\phi,\chi) =   {\lambda\over 4} \phi^4
+ {\alpha\over 4}\Bigl(\chi^2  - {M^2\over \alpha}\Bigr)^2 + {
1\over2} g^2 \phi^2 \chi^2 \ .
\end{equation}
 The models of such type have been  studied
in (Kofman and  Linde,  1987, Linde, 1991, 1994). We will assume here that
$\lambda \ll \alpha, g^2$, so
that
at large $\phi$ the curvature of the potential in the
$\chi$-direction is much
greater than in the $\phi$-direction. In this case at large $\phi$
the field
$\chi$ rapidly rolls toward  $\chi = 0$.
 An interesting feature of such models is the symmetry
restoration for the field $\chi$ for $\phi > \phi_c = M/g$, and
symmetry
breaking when the inflaton field $\phi$ becomes smaller than
$\phi_c$. As was
emphasized in (Kofman and  Linde,  1987), such phase transitions may lead to
formation
of
topological defects without any need for high-temperature effects.

We would like to point out some other specific features of such
models. If
the phase transition discussed above happens during inflation
(Kofman and  Linde,  1987)  (i.e.
if $\phi_c >
M_p$ in our model), then no new phase transitions occur in this model
after
reheating.
However, for $\phi_c \ll M_p$ the situation is much more complicated.
First of
all, in this case the field $\phi$ oscillates with the initial
amplitude $\sim M_p$ (if $M^4 < \alpha \lambda M_p^4$). This means
that each time when the absolute value of the
field becomes smaller than $\phi_c$, the phase transition with
symmetry
breaking occurs and topological defects are produced.
 Then the
absolute value
of the oscillating field $\phi$
again becomes greater than $\phi_c$, and symmetry restores again.
However, this regime does not continue for
a too long time. Within a few dozen  oscillations, quantum fluctuations of the
field
$\chi$ will be generated with the dispersion
$\langle(\delta\chi)^2\rangle \sim C^2 g^{-1}\sqrt\lambda M^2_{\rm P}
\ln^{-2}{1\over g^2}$ (Kofman {\it et al}, 1995, 1996).  For $M^2<C^2
g^{-1}\sqrt{\lambda}\alpha M_p^2\ln^{-2}{1\over g^2}$,
these fluctuations will keep the symmetry restored. Note that this effect may
be even stronger if instead of the term  ${\lambda\over 4} \phi^4$ we would
consider ${m^2\over 2} \phi^2$, since in that case the resonance is more broad
(Kofman {\it et al}, 1994). The symmetry breaking
finally completes  when $\langle(\delta\chi)^2\rangle$   becomes
small enough.

One may imagine even more complicated scenario when oscillations
of the
scalar field $\phi$ create large fluctuations of the field $\chi$,
which in
their turn interact with the scalar fields $\Phi$ breaking symmetry
in GUTs.
Then we would have phase transitions in GUTs induced by the
fluctuations of the
field $\chi$.  This means that no longer can the absence of primordial
monopoles be
considered as
an automatic consequence of inflation. To avoid the monopole production one
should  use the theories where   quantum fluctuations produced during
preheating are small or decoupled from the GUT sector. This condition imposes
additional constraints on realistic inflationary models. On the other hand,
preheating may remove some previously existing constraints on inflationary
theory.  For example, in the models of GUT baryogenesis it was assumed that
the GUT symmetry was restored by high temperature effects, since otherwise the
density of X, Y, and superheavy Higgs bosons would be very small.  This
condition is hardly compatible with inflation. It was also required that the
products of decay of  these particles should stay out of thermal equilibrium,
which is a very restrictive condition. In our case the superheavy particles
responsible for
baryogenesis can be abundantly produced by parametric resonance, and the
products of their decay will not be in a state of thermal equilibrium until the
end of reheating.

Now let us  return to   the theory (f{p1}) including the field
$\chi$ for
$g^2 \gg \lambda$. In this case the main fraction of the
potential energy density $\sim \lambda M^4_{\rm P}$ of the field $\phi$
predominantly
transfers to the energy of fluctuations of the field $\chi$
due to the explosive $\chi$-particles creation in the broad
parametric resonance.
The dispersion of fluctuations after preheating is
$\langle(\delta\chi)^2\rangle \sim C^2 g^{-1}\sqrt\lambda M^2_{\rm P}
\ln^{-2}{1\over g^2}$. These
fluctuations lead to the symmetry restoration
 in the theory (f{p1}) with
$\phi_0 \ll C \Bigl({g^2\over \lambda}\Bigr)^{1/4}  M_p\ln^{-1}{1\over g^2}$,
which may be much greater than $10^{16}$ GeV for $g^2 \gg \lambda$.

Later the process of decay of the field $\phi$ continues, but, just
as in the
model described in the previous section, one may
say with a good accuracy that
the fluctuations $\langle(\delta\chi)^2\rangle$ decrease as
$g^{-1}\sqrt\lambda M_p^2\,
 ({a_i\over a(t)})^2$ and their energy density $\rho$  decreases as
the energy
density of ultrarelativistic matter,  $\rho(t)  \sim \lambda M_p^4\,
({a_i\over a(t)})^{4}$, where $a_i$ is the scale factor at the end of
inflation. This energy density becomes equal to the vacuum energy
density
${m^4\over 4\lambda}$ at $a_0 \sim a_i \, \sqrt\lambda M_p/m,~
t\sim \sqrt{\lambda}M_pm^{-2}$. Since
that time
and until the time of the phase transition with symmetry breaking the
vacuum
energy dominates, and the universe enters secondary stage of  inflation.

The phase transition with spontaneous symmetry breaking occurs when
$m_{\phi ,eff}=0,~\langle(\delta\chi)^2\rangle = g^{-2} m^2$.
This happens at $a_c = a_i\,
\lambda^{1/4}g^{1/2} M_p/m$. Thus, during this additional
period of inflation
the universe expands ${a_c\over  a_0} \sim \sqrt g\,
\sqrt{\phi_0/m}=(g^2/\lambda)^{1/4}$ times.
This is   greater  than expansion during thermal inflation
(f{p5a}) by the
factor $O(g^{-1/2})$, and in our case inflation
occurs even if $g^4 \ll \lambda$.

In this example we considered the second stage of inflation driven by
the
inflaton field $\phi$. However, the same effect can occur in
theories where
other scalar fields are coupled to the field $\chi$. For example, in
the
theories of the type of  (f{p5b}) fluctuations
$\langle(\delta\chi)^2\rangle$
produced at the first stage of reheating by the oscillating inflaton
field
$\phi$ lead to a secondary inflation driven by the potential energy
of the
``flaton'' field $\Phi$. During this stage the universe expands $\sim
\sqrt g\,
 \sqrt{\Phi_0/m}$ times. To have a long enough inflation one may
consider,
e.g., supersymmetric theories with   $m\sim 10^2$ GeV and $\Phi_0
\sim 10^{12}$
(Lyth  and  Stewart, 1995). This gives a relatively long stage of inflation
with
${a_c\over
a_0} \sim \sqrt g \ 10^5$, which may be enough to solve the Polonyi
field
problem if the constant $g$ is not too small.

If the coupling constant $g$ is sufficiently
large,
fluctuations of the field $\chi$ will thermalize during this
inflationary
stage. Then the end of this stage will be determined by the standard
theory of
high temperature phase transition, and the degree of expansion during
this
stage will be given by $10^{-1} g\,  \sqrt{\Phi_0/m}$, see eq. (f{p5a}). It is
important,
however, that the inflationary stage may begin even if the field $\chi$
has not been thermalized at that time.

 The stage of inflation described above occurs  in the theory with a
potential which
is not particularly flat  near the origin. But what   happens in the
models
which have flat potentials, like the original new inflation model in
the
Coleman-Weinberg theory?
One of the main problems of inflation in such models was to
understand why
should the scalar field $\phi$ jump onto the top of its effective
potential,
since this field in realistic inflationary model is extremely weakly
interacting and, therefore, it could not be in the state of thermal
equilibrium in
the very early universe. Thus, it is much more natural for inflation
in the
Coleman-Weinberg theory to begin at very large $\phi$, as in the
simplest
version of chaotic inflation in the theory $\lambda \phi^4$.   However,
during the first few oscillations of the scalar
field $\phi$ at the end of inflation
in this model, it produces large non-thermal perturbations of vector
fields
$\langle (\delta A_\mu)^2 \rangle \sim C^2 g^{-1}\sqrt\lambda M^2_{\rm P}
\ln^{-2}{1\over g^2}$.  This
leads to
symmetry restoration  and initiates the second stage of inflation
beginning at $\phi = 0$.
It suggests that in many models inflation most naturally begins at
large
$\phi$ as in the simplest version of the chaotic inflation scenario. But then,
after the stage of preheating, the
second
stage of inflation may begin like in the new inflationary scenario.
Thus, the non-thermal symmetry restoration after chaotic inflation may
produce initial conditions which are necessary for
new inflation.

\section{Discussion}
Development of inflationary cosmology demonstrates over and over again that it
is dangerous to be dogmatic. For many years we believed that if observers  find
that $\Omega = 1$, they will prove inflation, and they will kill inflation if
they find that $\Omega$ differs from $1$ by more than about $10^{-4}$. This
made inflation  an easy and popular target for observers. Now we have found
that there exist several rather simple  models of an open inflationary
universe, according to which our universe consists of infinitely many domains
with all possible values of $\Omega$. This result is very encouraging for
theorists and somewhat disappointing for observers. Indeed, at the first glance
the measurement of $\Omega$ looses its fundamental importance, and  inflation
becomes a theory which is very difficult to verify. My opinion  is quite
opposite:  we have a win-win situation. If we   find   that $\Omega = 1$, it
will prove inflationary cosmology since 99\% of inflationary models predict
$\Omega = 1$, and no other theory makes this prediction. On the other hand, if
we find that $\Omega \not =1$, it will not disprove inflation, since now we
have inflationary models with $\Omega \not = 1$, and no other models of {\it
homogeneous and isotropic} universe with $\Omega \not = 1$ are known to us so
far.  Thus, inflationary theory becomes as robust as the whole Big Bang theory,
and it has a very nice property: It is possible to prove inflation, and it is
very hard to kill it.

On the other hand, until now we believed that inflation automatically solves
the primordial monopole problem. We thought that the physical processes after
inflation can be well understood as soon as we calculate the value of reheating
temperature. We have found that the situation is much more complicated, and,
consequently, much more interesting. In
addition to the standard high temperature phase transition, there exists
a  new class of phase transitions which
 may occur
 at the intermediate
stage between the end of inflation and the
establishing of thermal
equilibrium. These phase transitions may
 take place  even if the scale of symmetry breaking is very large and the
reheating temperature is very small. An important feature of these new phase
transitions is their non-universality. Indeed, they occur out of the state of
thermal equilibrium. Large quantum fluctuations are generated only for some
bose fields interacting with the inflaton field. As a result, it becomes
possible to have phase transitions producing superheavy strings, but to avoid
the phase transitions producing monopoles.  These phase transitions may lead to
    an efficient GUT baryogenesis,  and to existence of a
secondary stage of  inflation after  reheating.
 Therefore,
phase transitions of the new
type may
have dramatic consequences for inflationary models and the theory
 of physical processes in
the very
early universe.

\

\noindent{\bf Acknowledgements}
\medskip
\nobreak

It is a pleasure to thank Katsuhiko Sato and all organizers of the
Symposium for their  hospitality at the University of Tokyo.
I especially benefited from discussions with G. Efstathiou,  A. Vilenkin,
M. Rees, M. Sasaki and S. White.  This research was
supported in part by NSF
grant PHY-8612280.
\medskip
\medskip

\section*{\bf References}
\parskip 5pt plus 1pt

{}~~~~~1.  Abbott, L.F., Fahri, E., and  Wise, M., 1982, {\it Phys. Lett.} {\bf
117B}, 29.

 2.   Boyanovsky, D.,  de Vega, H.J,   Holman, R.,  Lee, D.S, and
 Singh, A., 1995,

{\it Phys.Rev. D} {\bf 51}, 4419.

  3.   Bucher, M. , Goldhaber, A.S., and Turok, N., 1995,  {\it Phys. Rev.}
D{\bf 52}, 3314.

  4. Bucher, M.,  and Turok, N., 1995,  {\it Phys. Rev.} D{\bf 52}, 5538.

  5. Coleman, S.,  and De Luccia, F.,  1980, {\it  Phys. Rev.} D {\bf 21},
3305
(1980).

 6. Dolan, L. and  Jackiw, R., 1974, {\it Phys. Rev.} {\bf D9}, 3357.

  7. Dolgov, A.D.  and Linde, A.D., 1982, {\it Phys. Lett.} {\bf 116B},  329.

  8. Dolgov, A.D. and  Kirilova, D.P.,  1990, {\it Sov. Nucl. Phys.}
{\bf 51},  273.

   9. Freese, K., Frieman, J.A., and  Olinto, A.V., 1990,
 {\it  Phys.\   Rev. \ Lett.} {\bf 65},  3233.

  10. Fujisaki, H.,  Kumekawa, K.,  Yamaguci, M., and Yoshimura, M., 1995,

Tokyo University preprints TU/95/488, hep-ph/9508378, and TU-95-493,

hep-ph/9511381.

   11. Futamase, T.  and Maeda, K.,  1989,  {\it Phys. Rev.} D {\bf 39},
399.

  12. Garc\'{\i}a--Bellido, J., 1995,  preprint SUSSEX-AST-95-10-1,
astro-ph/9510029.

  13. Garc\'{\i}a--Bellido, J., Linde, A.D. and Linde, D.A., 1994, {\it Phys.
Rev.}
{\bf D50}, 730.

   14. Garc\'{\i}a--Bellido, J. and Linde, A.D., 1995, {\it Phys. Rev.} {\bf
D51},
429.

  15. Gott, J.R., 1982,  {\it Nature} {\bf 295},   304  (1982).

  16. Hamazaki, T.,  Sasaki, M.,   Tanaka, T., Yamamoto, K., 1995,

 preprint  KUNS-1340,  gr-qc/9507006.

 17. Kaiser, D., 1995, Harvard University preprint HUTP-95/A027,
astro-ph/9507108.

 18. Kirzhnits, D.A., {\it JETP Lett.} {\bf 15}, 529 (1972).

 19. Kirzhnits, D.A. and  Linde, A.D., 1972, {\it Phys. Lett.} {\bf 42B}, 471.

 20. Kirzhnits, D.A. and  Linde, A.D., 1974, {\it ZhETF} {\bf 67}, 1263 ({\it
JETP} {\bf 40}, 628 (1975)).

 21. Kirzhnits, D.A. and  Linde, A.D., 1976, {\it Ann. Phys.} {\bf 101}, 195.

 22. Kofman, L.A. and  Linde, A.D.,  1987,  {\it Nucl. Phys.} {\bf B282},
555.

 23. Kofman, L.A.,    Linde, A.D. and
Starobinsky,  A.A., 1994, {\it Phys. Rev. Lett.}  {\bf 73}, 3195.

 24. Kofman, L.A.,    Linde, A.D. and
Starobinsky,  A.A., 1995, Stanford University

 preprint SU-ITP-95-21,

 hep-th/9510119, to be published in  {\it Phys. Rev. Lett.}.

 25. Kofman, L.A.,    Linde, A.D. and
Starobinsky,  A.A., 1996, in preparation.

  26.  Linde, A.D., 1984, {\it
Lett. Nuovo Cim.} {\bf 39}, 401

   27. Linde, A.D., 1986, {\it Phys. Lett.} {\bf B175}, 395.

  28.  Linde, A.D., 1990,   {\it Particle Physics and Inflationary
Cosmology}

 (Harwood, Chur, Switzerland).

  29.  Linde, A.D., 1991, {\it Phys. Lett.} {\bf B259}, 38.

  30.  Linde, A.D., 1994, {\it Phys. Rev.} {\bf D49}, 748.

  31.  Linde, A.D.,  1995 {\it  Phys. Lett.}  {\bf B351}, 99.

  32.  Linde, A.D., Linde, D.A. and Mezhlumian, A., 1994, {\it
Phys. Rev.} {\bf D49}, 1783.

  33.  Linde, A.D., Linde, D.A. and Mezhlumian, A., 1995a, Stanford University

 preprint  SU-ITP-95-25.

  34.  Linde, A.D.,  Mezhlumian, A.,  1995, {\it Phys. Rev.} D{\bf 52}, 6789.

 35.  Linde, A.D.,  Mezhlumian, A.,  1995a,  Stanford University preprint
SU-ITP-95-24,

 gr-qc/9511058, submitted
to Phys. Rev. D.

 36.   Linde, A.D., Linde, D.A. and Mezhlumian, A., 1995, {\it
Phys. Lett.} {\bf B345}, 203.

  37.  Lyth, D.H.  and  Stewart, E.D., 1995, {\it Phys. Rev. Lett.} {\bf 75},
201;  preprint

 LANCASTER-TH/9505, hep-ph/9510204.

 38.  Mukhanov, V.F.  and  Zelnikov, M.I.,  1991, {\it  Phys. Lett.}  {\bf
B263}, 169.

 39.  Sasaki, M., Tanaka, T., Yamamoto, K., and Yokoyama, J.,  1993, {\it
Phys. Lett.}
{\bf B317}, 510.

  40. Sasaki, M.,   Tanaka, T., Yamamoto, K., 1995, preprint  KUNS-1309,
astro-ph/9501109,

 to be published in ApJ.

  41. Sasaki, M., Tanaka, T., and Yamamoto, K., 1995, {\it Phys. Rev.} D {\bf
51}, 2979.

 42. Shtanov, Y.,   Traschen, J. and   Brandenberger, R., 1995, {\it
Phys.Rev.} D {\bf 51}, 5438.

  43. Starobinsky, A.A., 1984, in: {\it Quantum Gravity, Proc.
of the Second Seminar

 ``Quantum Theory  of Gravity'' (Moscow, 13-15
Oct. 1981)}, eds. M.A.  Markov and P.C. West

 (Plenum, New York), p. 103.

  44. Tanaka, T. and Sasaki, M., 1994, {\it Phys. Rev.} D{\bf 50}, 6444.

 45. Tkachev, I., 1995, Ohio State University preprint OSU-TA-21-95,
hep-th/9510146.

  46. Traschen, J., and  Brandenberger, R.,  1990, {\it  Phys. Rev.} D
{\bf 42},  2491.

  47. Vilenkin, A., 1984, {\it Phys. Rev.} D {\bf 30},   549.

 48. Vilenkin, A., 1995, {\it Phys. Rev.} {\bf D52}, 3365.

 49. Vilenkin, A. and Shellard, E.P.S., 1994,
{\it Cosmic Strings and Other
Topological Defects},

 Cambridge University Press, Cambridge.

 50. Weinberg, S., 1974, {\it Phys. Rev.} {\bf D9}, 3320.

  51. Winitzki, S. and Vilenkin, A., 1995,  gr-qc/9510054.

  52.  Yamamoto, K.,   Tanaka, T.,  and Sasaki, M.,1995, {\it Phys. Rev.} D
{\bf 51}, 2968.

 53. Yoshimura, M.,1995,  Tokyo University preprint TU/95/484 (1995),
hep-th/9506176.

\end{document}